\documentclass[amsmath,preprintnumbers,nofootinbib,final,5p,times,twocolumn]{revtex4-1}

\let\counterwithin\relax
\usepackage{amssymb}
\usepackage{amsmath}
\usepackage{color}
\usepackage{graphicx}
\usepackage{hyperref}
\usepackage{graphicx}
\usepackage{chngcntr}
\usepackage{float}
\usepackage[normalem]{ulem}
\graphicspath{{figures/}}

\hypersetup{
colorlinks = true,
linkcolor = blue,
anchorcolor = blue,
citecolor = blue,
filecolor = blue,
urlcolor = blue
}

\begin{document}

\title{Jet charge modification in finite QCD matter }

\author{Hai Tao Li }
\email{haitaoli@lanl.gov}

\author{Ivan Vitev}
\email{ivitev@lanl.gov}
\affiliation{Theoretical Division, Los Alamos National Laboratory, Los Alamos, New Mexico, 87545, USA}

\begin{abstract}
 Jet production and jet substructure modification in heavy-ion collisions have played an essential role in revealing the in-medium evolution of parton showers and the determination of the properties of  strongly interacting matter under extreme conditions.  It is imperative to extend these studies to include flavor tagging and to devise observables that are sensitive to the partonic origin of jets. The average jet charge,  defined as the momentum-weighted sum of the electric charges of particles inside the jet, is a proxy of the electric charge of the quark or gluon that initiates the jet. We demonstrate how the factorization framework of soft-collinear effective theory can be generalized to evaluate the jet charge in a dense strongly interacting matter environment,  such as the one produced in nuclear reactions at collider energies. Observables that can separate the contribution of in-medium branching from the trivial isospin effects are identified and their connection to established jet quenching effects is elucidated.  We present predictions for the transverse momentum dependence of the jet charge distribution in nucleus-nucleus collisions and its modification relative to the proton case.
\end{abstract}

\preprint{LA-UR-19-30442}

\maketitle

\section{Introduction}

Jet  production  in hadronic collisions is a ubiquitous and well-studied process in quantum chromodynamics 
(QCD)~\cite{Sterman:1977wj}. A new level of precision in the calculation of jet observables and insights into the substructure of jets has been achieved using the techniques of soft-collinear effective theory (SCET)~\cite{Bauer:2000ew,Bauer:2000yr,Bauer:2001ct,Bauer:2001yt,Beneke:2002ph}. In collisions of heavy nuclei, 
the cross section and substructure of jets
are modified by the formation of parton showers qualitatively different than the ones in the vacuum~\cite{Vitev:2008rz}.  These phenomena, generally known
as ``jet quenching",   provide a promising  avenue to study the properties of the quark-gluon plasma produced in heavy-ion collisions. 
%Observables related to jets has been widely used to extract the initial state nuclear effects and final state energy loss mechanism.  
In the past decade,  tremendous effort has been devoted to the experimental measurements and theoretical descriptions of jet production and jet properties
in such reactions.  The time is ripe for more differential studies that single out the production of jets of particular flavor. Some level of discrimination between inclusive jets and quark jets can be achieved via away-side photon tagging, which helps isolate the inverse Compton scattering process in QCD. Initial studies focused on the momentum imbalance distribution of vector boson-tagged jets in nucleus-nucleus (A+A) relative to the proton-proton $(p+p)$ 
collisions~\cite{Neufeld:2012df,Dai:2012am,Qin:2012gp,Wang:2013cia,Chatrchyan:2012gt,Sirunyan:2017qhf,Sirunyan:2017jic,Aaboud:2018anc}. 
More recently, the jet substructure modification of  photon-tagged predominantly quark jets has been compared to the corresponding modification of 
inclusive jets~\cite{Chien:2015hda,Sirunyan:2018ncy,Aaboud:2019oac}.  Ultimately, we would like to understand the modification of individual flavor jets, 
such as up-quark jets or down-quark jets. 
  
Up quarks, down quarks and gluons carry different electric charge. Even though the electric charge of a quark or a  gluon cannot be directly measured, it  can be estimated from the charge of jets initiated by the corresponding hard partons. The jet charge  is defined as the transverse momentum-weighted sum of the charges of the jet constituents~\cite{Field:1977fa}  
\begin{align} \label{eq:charge}
    Q_{\kappa, {\rm jet}}  = \frac{1}{\left(p_T^{\rm jet}\right)^\kappa } \sum_{\rm i\in {\rm jet}} Q_i \left(p_T^{i} \right)^{\kappa } \;  ,
\end{align}
where the sum runs over  all particles in the reconstructed jet with transverse momentum $p_T^{\rm jet}$. $Q_i$ and $p_T^{i}$ are the electric charge  and the transverse momentum, respectively, of  particle $i$. Here,  $\kappa$ is a free parameter with the requirement that $\kappa > 0$. We will later show that it can be chosen to enhance the sensitivity to medium-induced parton shower effects for the individual flavor jet charge. 

Jet charge measurements date back to the late 1970s and the early 1980s~\cite{Berge:1980dx,Albanese:1984nv,Erickson:1979wa}.  
This observable has found a variety of  applications, such as identifying the charge of b-quark jets~\cite{ Braunschweig:1990cv, Abreu:1991yf, Decamp:1991se, Acton:1992sc,Abreu:1996hz,Abe:1994bm,Abe:1999ds}, and the $W$-boson charge~\cite{Barate:1997ts,Abreu:2001rpa,Acciarri:1999kn,Abbiendi:2000ej} for a wide array of Standard Model measurements.
Using dijet events, the jet charge distribution has been measured at the Large Hadron Collider (LHC) by the ATLAS and CMS Collaborations~\cite{TheATLAScollaboration:2015bgc,Aad:2015cua, CMS:2016yuu,Sirunyan:2017tyr}.
In particular, the ATLAS measurements reported in Ref.~\cite{Aad:2015cua} extract the average up-quark and down-quark jet charges as a function of jet $p_T$. The positive (up-quark jet) and negative (down-quark jet) electrically charged jets can be clearly distinguished, and the measurements also confirm the scale violation of the quark jet charge predicted by Refs.~\cite{Krohn:2012fg,Waalewijn:2012sv},  which reads
\begin{align} \label{eq:scalev}
    \frac{p_T}{\langle Q_{\kappa, q}  \rangle} \frac{d}{dp_T}\langle Q_{\kappa, q} \rangle = \frac{\alpha_s}{\pi} \tilde{P}_{qq}(\kappa),
\end{align}
where $ \tilde{P}_{qq}(\kappa)$ is the $(\kappa+1)$th Mellin moment of the leading-order splitting function.  This is the main motivation to extend this
observable to heavy-ion collisions. 

In the SCET framework  it was found~\cite{Krohn:2012fg,Waalewijn:2012sv}  that the jet charge can be written as the product of the jet matching coefficients and the nonperturbative fragmentation function in proton-proton collisions.\footnote{ For a detailed discussion of the jet matching coefficients and jet functions, see Ref.~\cite{Ellis:2010rwa}. Definitions of Mellin moments relevant to the evaluation of the jet charge are given in Sec.~\ref{sec:jetcharge}. We briefly present the calculation of the jet matching coefficients and functions in Appendix~\ref{app:JetF}.}
Furthermore, for narrow and well-separated energetic jets, the jet charge is independent of the hard process.  Assuming that soft correlations are negligible, the gluon jets always give zero jet charge because the  contributions from quarks and antiquarks generated by gluon splitting cancel out.  The jet charge can be used to separate the quark jets from antiquark jets and to distinguish the quark flavor, as a recent study~\cite{Fraser:2018ieu} using modern machine-learning techniques has shown. Note that jet charge has to be defined at the level of hadrons and hadronization effects must be taken into account.

The jet charge distribution is  a particularly interesting and, in fact, complex observable.  Measurements of jet charge can enhance our understanding of nuclear modification in heavy-ion collisions, including initial-state and   final-state effects. 
First, the fractions of up-quark and down-quark jets in A+A collisions are significantly modified when compared to the ones in $p+p$ collisions due to isospin effects. The jet charge distribution is very sensitive to the flavor properties, which can be used to constrain the global-fit nuclear parton distribution functions (PDFs),  for example nCTEQ15 PDF sets~\cite{Kovarik:2015cma}. While isospin effects are rather trivial, there are not many ways to accurately test them. The quark flavor
composition and electric charges affect the cross sections for direct photon production~\cite{Vitev:2008vk,Adare:2012vn,Aaboud:2019tab,David:2019wpt}.  Precise knowledge of these effects is essential to uncover initial-state inelastic processes in cold nuclear matter that can  further modify the cross sections for
particle and jet production and manifest themselves in correlations between the soft particles produced in the collision event~\cite{Albacete:2017qng,Gyulassy:2014cfa}.

Second, medium-induced parton showers affect the propagation of quark jets and gluon jets in dense QCD matter differently because of the different color charges. The stronger suppression of gluon-initiated jets reduces the ``dilution" of the jet charge, as gluon jets carry zero average charge. In heavy-ion collisions this differential quenching can affect other jet substructure observables as well~\cite{Chien:2015hda}. 

Finally, the evolution of the jet function and fragmentation functions is also modified in the QCD medium~\cite{Majumder:2011uk,Kang:2014xsa,Chien:2015vja}. This is certainly  the most interesting  effect in the modification of the jet charge. Given the fraction of different types of jets and the measurements of the jet charge distribution, as was done in Ref.~\cite{Aad:2015cua} using different rapidities, the scale violation parameter of the up-quark or down-quark jet can be extracted.  It is determined by the $(\kappa+1)$th Mellin moment of the medium-induced splitting function shown in Eq.~(\ref{eq:scalev}). This will provide a unique and nontrivial test on the evolution of energetic partons in a QCD medium.

With this in mind, we embark on a theoretical study of the jet charge in heavy-ion collisions. We note that very recently simulations of the jet 
charge in nucleus-nucleus reactions  were presented in Ref.~\cite{Chen:2019gqo}, fully relying on Monte Carlo event generators.  In this work our goals are 
somewhat different  --  we  present a framework for perturbative calculations of the jet charge   in heavy-ion collisions  and its modification relative
 to proton collisions building upon the approach developed in Refs.~\cite{Krohn:2012fg,Waalewijn:2012sv}. 
  In analogy to the vacuum case, the medium corrections to the jet function are constructed  with the help of the medium-induced
splitting functions~\cite{Ovanesyan:2011xy,Ovanesyan:2011kn,Kang:2016ofv,Sievert:2018imd,Sievert:2019cwq} that capture the full collinear branching dynamics 
of energetic parton evolution in a QCD medium. A similar application of the medium-induced splitting functions can be found in Refs.~\cite{Kang:2017frl,Li:2018xuv}. In addition to the jet function, in heavy-ion collisions the evolution of the  fragmentation functions is controlled by the full splitting function $P_{ij} \to P_{ij}+P_{ij}^{\rm med}$~\cite{Kang:2014xsa,Chien:2015vja}. As mentioned earlier, initial-state effects are considered using the global-fit nuclear PDFs.  Finally, 
we discuss how the medium-induced shower evolution effects on the jet charge can be disentangled from the more trivial 
isospin and jet quenching effects.

In the following section we review the calculation of the average jet charge in proton-proton collisions in the framework of SCET. The extension to the heavy-ion collisions is presented in Sec.~\ref{sec:HIC} using the in-medium splitting functions. Numerical results are contained in Sec.~\ref{sec:numerical}. We conclude in Sec.~\ref{sec:concl}. A derivation of the jet function and the calculated fraction of different flavors of jets are enclosed in the Appendices.

\section{The jet charge} \label{sec:jetcharge}
Following Refs.~\cite{Krohn:2012fg,Waalewijn:2012sv} we first briefly review the calculation of the average jet charge. 
The average charge for gluon jets is zero if  soft correlations are ignored, and we will work in this approximation. The average  charge of  a quark jet (q jet) is 
given by
\begin{align}
    \langle Q_{\kappa, q} \rangle = \int dz~z^\kappa \sum_{h} Q_h \frac{1}{\sigma_{\text{q jet}}} \frac{d\sigma_{h\in \text{q jet}}}{dz} \;. 
\end{align}
Here, $Q_h$ is the charge of the hadron inside the jet and $z= p_T^{i}/p_T$ is the corresponding momentum fraction.
From the factorization of jet production in SCET~\cite{Procura:2009vm,Liu:2010ng, Ellis:2009wj, Ellis:2010rwa, Jain:2011xz,Chien:2015ctp,Dai:2016hzf} the average charge for q jet can be written as 
\begin{align} \label{eq:Q}
     \langle Q_{\kappa, q} \rangle = \frac{\tilde{\mathcal{J}}_{qq}(E,R,\kappa,\mu)}{J_{q}(E,R,\mu)} \tilde{D}_q^{Q}(\kappa,\mu) \;, 
\end{align}
where $J_{q}(E,R,\mu)$ is a jet function and $\tilde{\mathcal{J}}_{qq}(E,R,\kappa,\mu)$ is the Wilson coefficient for  matching  the quark  fragmenting  jet  function  onto  a quark fragmentation  function. In Eq.~(\ref{eq:Q}) $E$ is the jet energy,  $R$ is the jet radius, and $\mu$ is the factorization scale.
The $(\kappa+1)$th Mellin moments of the jet matching coefficient and fragmentation function are defined as
\begin{align}
    \tilde{\mathcal{J}}_{qq}(E,R,\kappa,\mu) &= \int_0^1 dz~ z^{\kappa} \mathcal{J}_{qq}(E,R,z,\mu)  \;, 
    \nonumber \\ 
    \tilde{D}_q^{Q}(\kappa,\mu) &= \int_0^1 dz ~z^{\kappa} \sum_{h} Q_h D_q^{h}(z,\mu) \;.
\end{align}
The perturbative next-to-leading-order  (NLO) $k_T$-like jet function and the matching coefficients from the jet to the hadron can be found in Refs.~\cite{Waalewijn:2012sv,Jain:2011xz,Ellis:2010rwa}. Using dimensional regularization, we derived the jet matching coefficients $\mathcal{J}_{qq}$ in Appendix~\ref{app:JetF}. The nonperturbative  fragmentation function $D_j^{h}$ describes the probability to produce hadron $h$ from a parton $j$ and obeys the timelike Dokshitzer-Gribov-Lipatov-Altarelli-Parisi (DGLAP) evolution equations. 
The evolution of the charge-weighted fragmentation function  $\tilde{D}_q^{Q}(\kappa,\mu)$  is then given by
\begin{align} \label{eq:Qevol}
   \mu \frac{d}{d\ \mu} \tilde{D}_q^{Q}(\kappa,\mu) = \frac{\alpha_s(\mu)}{\pi} \tilde{P}_{qq}(\kappa) \tilde{D}_q^{Q}(\kappa,\mu)  \;, 
\end{align}
where $\tilde{P}_{ij}(\kappa)$ is the $(\kappa+1)$th Mellin moment of the NLO splitting function $P_{ij}$ and
\begin{align}
    \tilde{P}_{qq}(\kappa) = C_F \int_0^1 dz~ (z^\kappa-1)\frac{1+z^2}{1-z} \;,
\end{align}
where $-1$ arises from the plus prescription.
For a given $\kappa$, the jet charge depends on only one nonperturbative parameter $\tilde{D}_q^{Q}(\kappa,\mu_0)$.  Notice that $\langle Q_\kappa \rangle$ is free of the scale $\mu$ up to the  perturbative order that we employ. 

\section{Jet charge modification in heavy-ion collisions}\label{sec:HIC}
In heavy-ion collisions,  jet production receives medium-induced modifications.  As discussed in the introduction  and seen in Eq.~(\ref{eq:Q}),  the jet charge is modified through the medium-induced corrections to the jet function, jet matching coefficient, and the fragmentation functions. The jet functions $J_i$ and  $\tilde{\mathcal{J}}_{ij}$ are constructed using the medium-induced splitting functions. The fragmentation functions and their evolution in Eq.~(\ref{eq:Qevol}) are also modified in the QCD medium.

\subsection{Medium-induced splitting functions}

Propagation of partons in QCD matter adds a medium-induced component to the parton showers that characterize simpler reactions, such as 
$e^+ + e^-$, $e+p$ and $p+p$. The in-medium branching processes relevant to shower formation can be calculated  order by order in powers of opacity, 
or the mean number of scatterings in the medium.  The Relation of this technique to other approaches to evaluate inelastic processes in matter is discussed in  Ref.~\cite{Sievert:2018imd}.  The opacity expansion  was first developed in the soft-gluon emission parton  energy loss limit.   At LHC energies early works~\cite{Gyulassy:2000fs,Gyulassy:2000er} found that the first order in opacity is a good approximation for jet quenching
applications. Recently, the full splitting functions have been calculated analytically to higher orders in opacity~\cite{Sievert:2018imd,Sievert:2019cwq}, but their 
numerical evaluation remains computationally intensive. The evaluation in higher orders in opacity is also difficult in the soft-gluon emission limit~\cite{Feal:2019xfl, Andres:2020vxs}. 
For these reasons  we will use the medium-induced splitting functions up to first order in opacity and  focus on light flavor jets. 
We note that the accuracy of theoretical predictions 
to this order has been confirmed by experimental measurements for both light and heavy hadron and jet observables~\cite{Chien:2015vja,Kang:2016ofv,Kang:2017frl,Li:2018xuv,Sirunyan:2018ncy}.   
The complete sets of the massless splitting kernels to that order can be found in Ref.~\cite{Ovanesyan:2011kn}.  
The real contribution to the splitting function can be written as 
\begin{align}
    P_{i \rightarrow j k }^{\mathrm{med},\rm{real}} \left(x, \mathbf{k}_{\perp}\right) = 2\pi \times \mathbf{k}_{\perp}^2  \frac{dN_{i \rightarrow j k }^{\mathrm{med}} }{d^2\mathbf{k}_{\perp}  dx} \,,
\end{align}
where $x$ is momentum fraction of parton $j$ from the parent parton $i$ and $\mathbf{k}_{\perp}$ is the transverse momentum of parton $j$ relative the parton $i$.  The reader will note that 
in this paper we follow the standard high-energy physics convention for the branching kinematics and soft gluon emission corresponds to the large-$x$ limit. 

As is the case with the vacuum splitting functions, the NLO full medium splitting functions include  contributions with and without a real emission.  This can be represented as a plus prescription regularized 
function in the limit  $x \to 1$  and a Dirac-delta function.  For the case of $q\to q g$ the full splitting functions is defined as  
\begin{align}\label{eq:pqq}
    P_{q \rightarrow q g}^{\rm{med}}\left(x, \mathbf{k}_{\perp}\right)=\left[P_{q \rightarrow q g}^{\mathrm{med},\rm{real}} \left(x, \mathbf{k}_{\perp}\right)\right]_{+}+A\left(\mathbf{k}_{\perp} \right) \delta(1-x) \; .
\end{align}
With the choice to regularize the full expression,  the  coefficient function function $A\left(\mathbf{k}_{\perp} \right)$ is obtained using the flavor conservation sum rule,
\begin{align} \label{eq:pqqFlavor}
    \int_0^1 dx  P_{q \rightarrow q g}^{\rm{med}}\left(x, \mathbf{k}_{\perp} \right) =A\left(\mathbf{k}_{\perp} \right)=  0 \;. 
\end{align}
The medium-induced splitting functions for the $q\to gq$ channel is  
\begin{align}
     P_{q \rightarrow  g q }^{\rm{med}}\left(x, \mathbf{k}_{\perp} \right)  
     &= P_{q \rightarrow g q}^{\mathrm{med},\rm{real}} \left(x, \mathbf{k}_{\perp}\right) = P_{q \rightarrow  qg}^{\mathrm{med},\rm{real}} \left(1-x, \mathbf{k}_{\perp}\right) \;  .
\end{align}
The splitting functions for gluon-initiated channels are not relevant to the jet charge calculations.  For the convenience of readers, 
the application of the full set of the medium splitting functions and relations among them can be found in Refs.~\cite{Kang:2014xsa,Chien:2015vja}. 

As mentioned in the  introduction, the medium-induced splitting kernels are the analog of the vacuum Altarelli-Parisi splitting kernels and in the presence of a medium
it is easy to show by writing down the relevant Feynman diagrams that for their continuous part $P_{ij}  \left(x\right )   \rightarrow P_{ij}  \left(x \right) +P_{ij}^{\rm med}  \left(x, \mathbf{k}_{\perp}\right)$~\cite{Ovanesyan:2011kn}.  The real part can analytically be written as a correction to Altarelli-Parisi    
$P_{ij}^{\rm real}  \left(x\right )   \rightarrow P_{ij}^{\rm real}  \left(x \right) \left[ 1  +g_{ij}^{\rm med,real}  \left(x, \mathbf{k}_{\perp}\right) \right ]$   and the virtual corrections for the diagonal branchings
can be computed from momentum and flavor sum rules.

\subsection{Medium modifications to the factorized jet charge calculation}

We explicitly derive the vacuum jet function and jet matching coefficient in Appendix~\ref{app:JetF}. In analogy to the case of the vacuum function shown in Eq.~(\ref{eq:jq2q}), the medium modifications are introduced by replacing the splitting kernels with the in-medium splitting functions.  In the medium splitting kernel, the pole when $x\to 1$ is regularized by the plus distribution function as shown in Eq.~(\ref{eq:pqq}).  Because we cannot use dimensional regularization to deal with the ultraviolet divergences in the medium sector, similar to the case of semi-inclusive jet functions~\cite{Kang:2017frl,Li:2018xuv}, the medium correction to  $ \mathcal{J}_{qq}$ is calculated using the medium-induced splitting kernel, 
\begin{align}
    \mathcal{J}_{qq}^{\rm med }(E,R,x,\mu) =&  
    \nonumber \\  & \hspace{-2cm}  
    \frac{\alpha_s (\mu)}{2\pi^2}  \left[  - \delta(1-x)\int_0^1 dz \int_0^{\mu}  \frac{d^2 \mathbf{k}_{\perp} }{\mathbf{k}_{\perp}^2}    P_{q \rightarrow q g}^{\rm{med}}\left(z, \mathbf{k}_{\perp}\right) 
   \right.  \nonumber \\  & \left.  \hspace{-2cm} +
     \int_0^{2 E x(1-x)\tan R/2} \frac{d^2 \mathbf{k}_{\perp} }{\mathbf{k}_{\perp}^2}  P_{q \rightarrow q g}^{\rm{med}}\left(x, \mathbf{k}_{\perp}\right)  \right] 
     \nonumber \\  & \hspace{-2cm}  
     =\frac{\alpha_s(\mu)}{2\pi^2}  \int_0^{2 E x(1-x)\tan R/2} \frac{d^2 \mathbf{k}_{\perp} }{\mathbf{k}_{\perp}^2}  P_{q \rightarrow q g}^{\rm{med}}\left(x, \mathbf{k}_{\perp}\right) \,,
\label{match}
\end{align}
where the second line is the virtual corrections and the third line is the real corrections. 
 The virtual corrections do not give a contribution to the matching coefficient and 
can be understood  from flavor conservation and the connection between the flavor and electric charge of jets of fixed flavor.  The medium correction to  $ \mathcal{J}_{qg}$ is 
\begin{align}    
    \mathcal{J}_{qg}^{\rm med }(E,R,x,\mu)  =&
    \nonumber \\ &  \hspace{-2cm}
    \frac{\alpha_s(\mu)}{2\pi^2} \int_0^{2 E x(1-x)\tan R/2} \frac{d^2 \mathbf{k}_{\perp} }{\mathbf{k}_{\perp}^2}  P_{q \rightarrow g q}^{\rm{med}}\left(x, \mathbf{k}_{\perp}\right) \,,
\end{align}
where the integrals are defined in four-dimensional spacetime. As in Eq.~(\ref{eq:totaljet}) the medium correction to the total quark jet function is 
\begin{widetext}
\begin{align}
    J_{q}^{\rm med}(E,R,\mu) & = 
    \int_0^1 dx~ x \bigg(\mathcal{J}_{qq}^{\rm med}(E,R,x,\mu)+\mathcal{J}_{qg}^{\rm med}(E,R,x,\mu)\bigg)   
    \nonumber \\ & = 
     \frac{\alpha_s(\mu)}{2\pi^2} \int_0^1 dx  \int_{0}^{2 E x(1-x)\tan R/2} \frac{d^2 \mathbf{k}_{\perp} }{\mathbf{k}_{\perp}^2 }
    \bigg(x P_{q \rightarrow  qg }^{\rm{med,real}}\left(x, \mathbf{k}_{\perp} \right) +x P_{q \rightarrow  gq }^{\rm{med,real}}\left(x, \mathbf{k}_{\perp} \right) \bigg)
    \nonumber \\ &  =  \frac{\alpha_s(\mu)}{2\pi^2} \int_0^1 dx  \int_{0}^{2 E x(1-x)\tan R/2} \frac{d^2 \mathbf{k}_{\perp} }{\mathbf{k}_{\perp}^2 }
    P_{q \rightarrow  qg }^{\rm{med,real}}\left(x, \mathbf{k}_{\perp} \right),
 \label{inclfunc}
\end{align} 
\end{widetext}
where  in the third line we used the fact that $ P_{q \rightarrow  gq }^{\rm{med,real}}(x, \mathbf{k}_{\perp}) =  P_{q \rightarrow  qg }^{\rm{med,real}}(1-x,\mathbf{k}_{\perp}) $.

To the order that we calculate, see Appendix~A, the $R$-dependent upper limit of the $ \mathbf{k}_{\perp}$ integration  in Eqs. (\ref{match}) - (\ref{inclfunc}) determines what part of the medium-induced parton shower enters the reconstructed jet~\cite{Kang:2017frl,Li:2018xuv}.  This, in turn, is reflected in the suppression of jet cross sections in heavy-ion 
versus proton collisions~\cite{Vitev:2008rz,Vitev:2009rd}.  Recent CMS measurements~\cite{CMS:2019btm} have shown that the SCET jet function approach gives an excellent description of the
radius dependence of jet cross sections in Pb+Pb reactions at the LHC. They are a nontrivial check of the formalism that we use here to evaluate the jet charge.

The collinear radiation in a QCD medium beyond leading order can be included through solving the medium-modified DGLAP equations. This technique has been used extensively in Refs.~\cite{Wang:2001ifa,Wang:2009qb,Majumder:2013re,Chang:2014fba,Kang:2014xsa,Chien:2015vja,Kang:2016ofv,Li:2017wwc} to describe hadron production and carry out resummation  numerically   in a strongly interacting environment. Details of the theoretical formalism we use  are given in~\cite{Kang:2014xsa,Chien:2015vja} and its predictions have
 been validated by inclusive hadron suppression measurements~\cite{Khachatryan:2016odn}.  In a QCD medium the evolution of the charge-weighted fragmentation function becomes
\begin{align} \label{eq:AAQevol}
  \frac{d}{d\ln \mu} \tilde{D}_q^{Q, {\rm full}}(\kappa,\mu)  =
        \nonumber \\ &  \hspace{-2.5cm}    
   \frac{\alpha_s(\mu)}{\pi} \left(\tilde{P}_{qq}(\kappa)+\tilde{P}^{\rm med}_{qq}(\kappa,\mu) \right) \tilde{D}_q^{Q,{\rm full}}(\kappa,\mu) ,
\end{align}
where $\tilde{P}^{\rm med}_{qq}(\kappa,\mu)$ is the $(\kappa+1)$th Mellin moment of the medium splitting kernel 
\begin{align}
    \tilde{P}^{\rm med}_{qq}(\kappa,\mu) =   \int_0^1 dx ~x^{\kappa}~ P_{q \rightarrow   qg }^{\rm{med}}\left(x, \mathbf{k}_{\perp} \right)\bigg|_{k_\perp=\mu}~.
\end{align}
 In the above equation, the typical scale of the collinear splitting in medium is set to be $\mu= k_\perp$~\cite{Kang:2014xsa, Chien:2015vja,Kang:2016ofv,Li:2017wwc}. 
 The additional scale dependence in the medium-induced part of Eq.~(\ref{eq:AAQevol}) reflects the difference in the $k_\perp$ dependence of the vacuum and in-medium 
 branching processes.
 
After combining all the medium-modified components of the expression together, the average jet charge in heavy-ion collisions reads
%\begin{widetext}
\begin{align} \label{eq:AAQ}
    \langle Q_{q, \kappa}^{\rm AA} \rangle = &
    \frac{\tilde{\mathcal{J}}_{qq}(E,R,\kappa,\mu)+\tilde{\mathcal{J}}^{\rm med}_{qq}(E,R,\kappa,\mu)}{J_{q}(E,R,\mu)+J_{q}^{\rm med}(E,R,\mu)}
    \nonumber \\ & \times 
    \tilde{D}_q^{Q,{\rm full}}(\kappa,\mu)~
    \nonumber \\  
    =&\langle Q_{q, \kappa}^{\rm pp} \rangle \left(1+\tilde{\mathcal{J}}^{\rm med}_{qq}-J_{q}^{\rm med}\right) 
    \nonumber \\ & \times 
    \exp\left[ \int_{\mu_0}^{\mu} \frac{d\bar{\mu}}{\bar{\mu}}    \frac{\alpha_s(\bar{\mu})}{\pi}  \tilde{P}^{\rm med}_{qq} \right] + \mathcal{O}(\alpha_s^2,~\chi^2)\,,
\end{align}
where we have expanded the medium-induced jet function and jet matching coefficient moment to first nontrivial order and   $\chi$ is the  opacity expansion parameter. 
For the medium correction to the jet function we obtain explicitly
\begin{multline}
    \tilde{\mathcal{J}}^{\rm med}_{qq}-J_{q}^{\rm med} = \frac{\alpha_s(\mu)}{2\pi^2} \int_0^1 dx~(x^\kappa-1)
    \\ \times
    \int_{0}^{2 E x(1-x)\tan R/2} \frac{d^2 \mathbf{k}_{\perp} }{\mathbf{k}_{\perp}^2 }
    P_{q \rightarrow  qg }^{\rm{med,real}}\left(x, \mathbf{k}_{\perp} \right)\,.
\end{multline}

Note that in the medium the scale $\mu$ dependence is not canceled after combining the jet function and fragmentation function. The scale choice is related to physical 
parameters, such as the transverse momentum $p_T$ of the jet and the jet radius $R$.  As a result of the Landau-Pomeranchuk-Migdal effect~\cite{Landau:1953um,Migdal:1956tc}  in QCD, medium-induced branchings
depend on such energy and resolution scales that are combined into the choice of $\mu$.
 Up to NLO in QCD and first order in opacity, the  scale dependence comes exclusively from the medium and can be written as 
\begin{align}
    \frac{d}{d\ln \mu} \langle Q_{q, \kappa}^{\rm AA} \rangle  =  \frac{\alpha_s(\mu)}{\pi} \tilde{P}^{\rm med}_{qq}(\kappa,\mu)   \langle Q_{q, \kappa}^{\rm AA}  \rangle \,,
\end{align}
where the medium splitting function $\tilde{P}^{\rm med}_{qq}(\kappa,\mu)$ is approximately zero for very high-energy jets. 
At the scale $\mu = p_T R $, the $p_T$ dependence of the jet charge is 
\begin{multline} \label{eq:mediumQT}
        \frac{d}{d\ln p_T} \ln \langle Q_{q, \kappa}^{\rm AA} \rangle  = 
        \frac{\alpha_s(p_T R)}{\pi} \left[  \tilde{P}_{qq}(\kappa)         
        + \tilde{P}^{\rm med}_{qq}(\kappa,p_T R) 
        \right. \\ \left.
        + \int_0^1 dx (x^\kappa-1) P^{\rm med}_{qq}(\kappa,{k}_\perp = x(1-x) p_T R)  \right]  \,,
\end{multline}
where $p_T R \approx 2 E \tan(R/2)$.

\begin{figure}[htb!]
    \centering
    \includegraphics[scale=0.47]{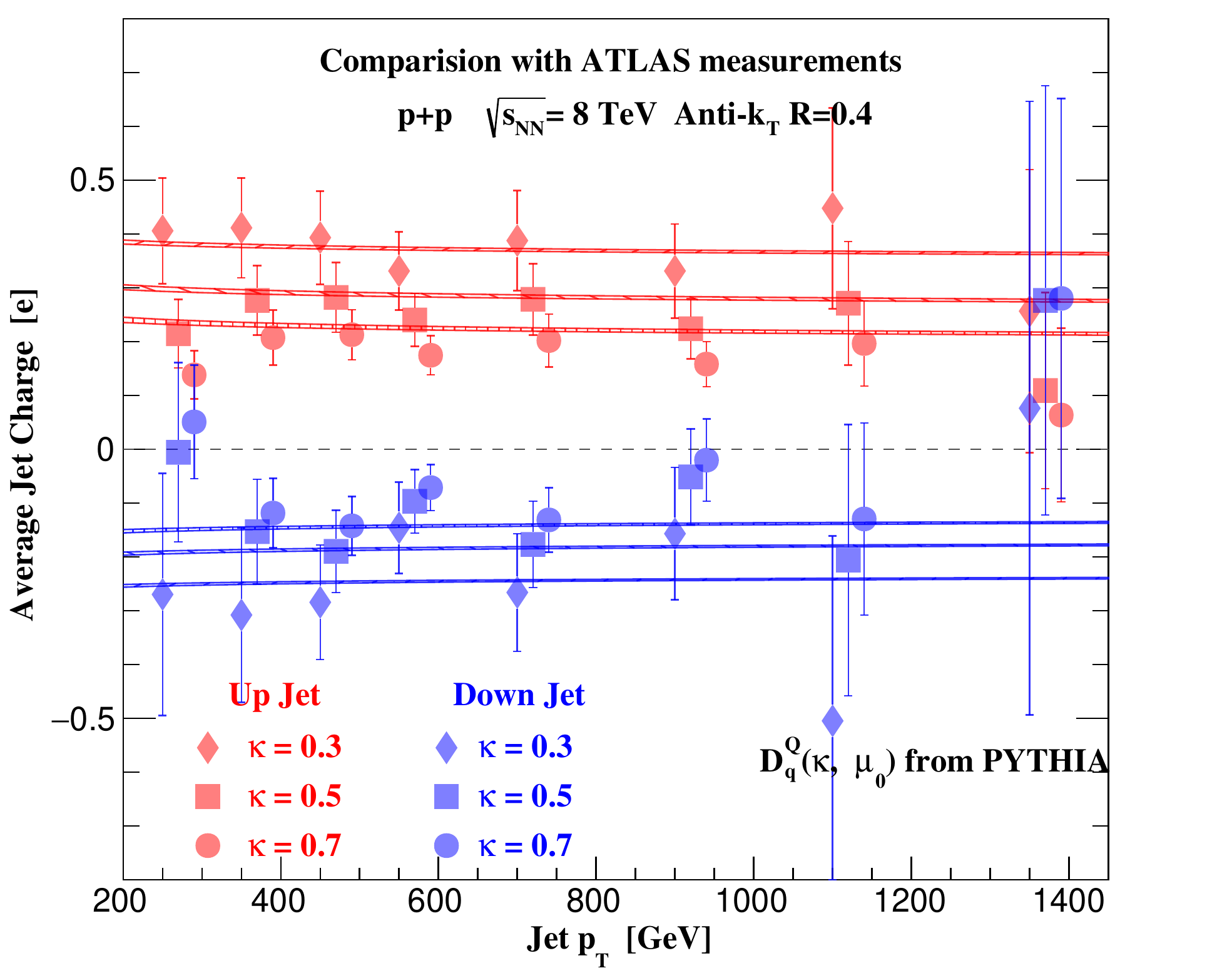}
    \vspace{-0.4cm}
    \caption{Average charge of up- and down-quark jets as a function of jet $p_T$ with $\kappa=$0.3, 0.5 and 0.7 in  $\sqrt{s_{\rm NN}} = 8$~TeV  $p+p$ collisions. The dots with error bars represent the measurements by ATLAS~\cite{Aad:2015cua}.  The red and blue lines are the prediction with the nonperturbative parameters of hadronization  obtained through fitting to PYTHIA simulations. }
    \label{fig:ppQ}
\end{figure}

\begin{figure*}[htb!]
    \centering
    \includegraphics[scale=0.45]{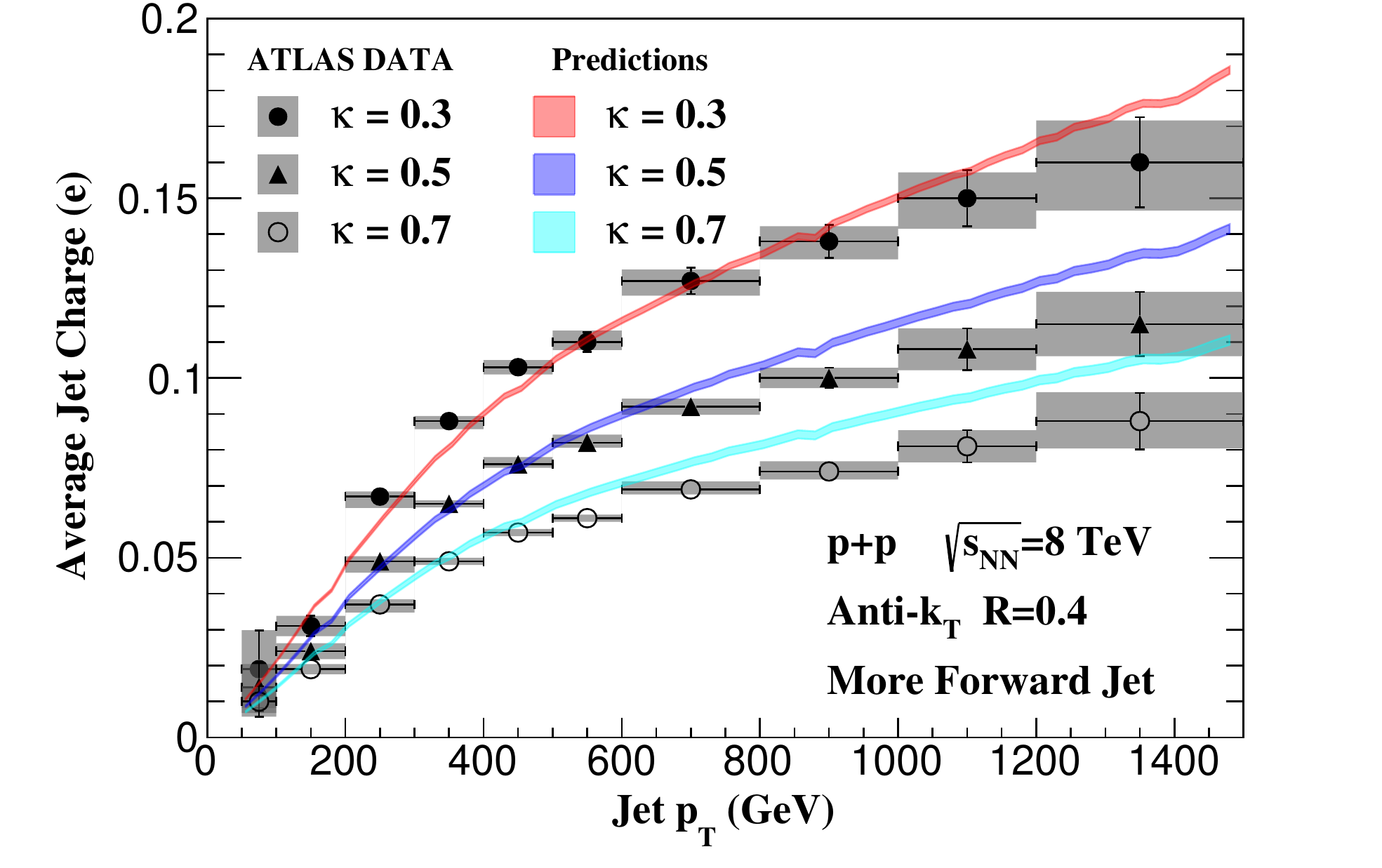} \hspace{-0.6cm}
    \includegraphics[scale=0.45]{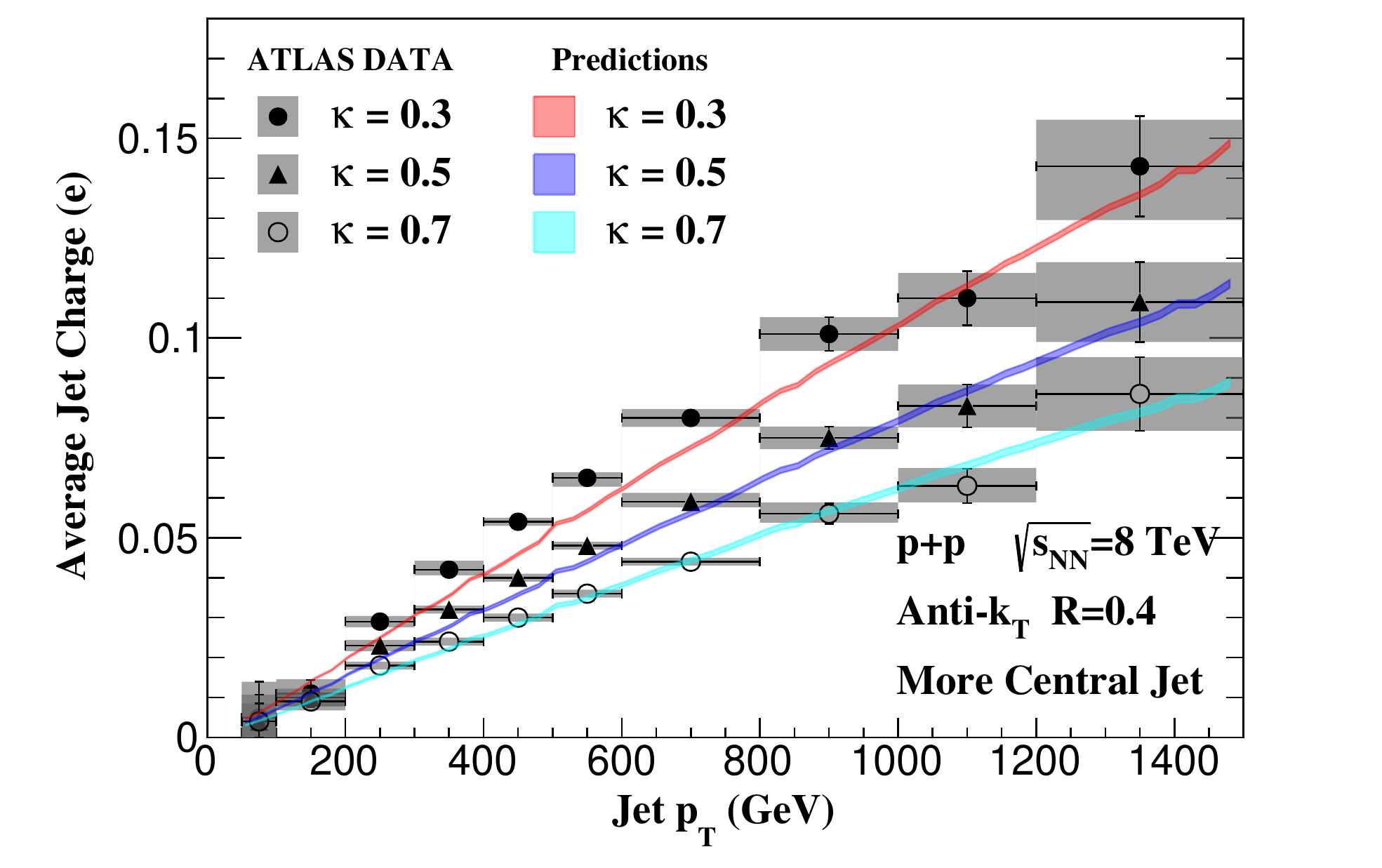}    
    \vspace{-0.4cm}    
    \caption{ Transverse momentum dependence of the average  jet charge distribution for the more forward jet (left) and more central jet (right) in dijet events at 
    $\sqrt{s_{\rm NN}} = 8$~TeV with  $\kappa=$0.3, 0.5 and 0.7 in $p+p$ collisions at the LHC. The gray bars represent the total experimental uncertainty of the data, while the colored bands are theoretical predictions. }
    \label{fig:QoverPT}
\end{figure*}

\begin{figure}[htb!]
    \centering
    \includegraphics[scale=0.45]{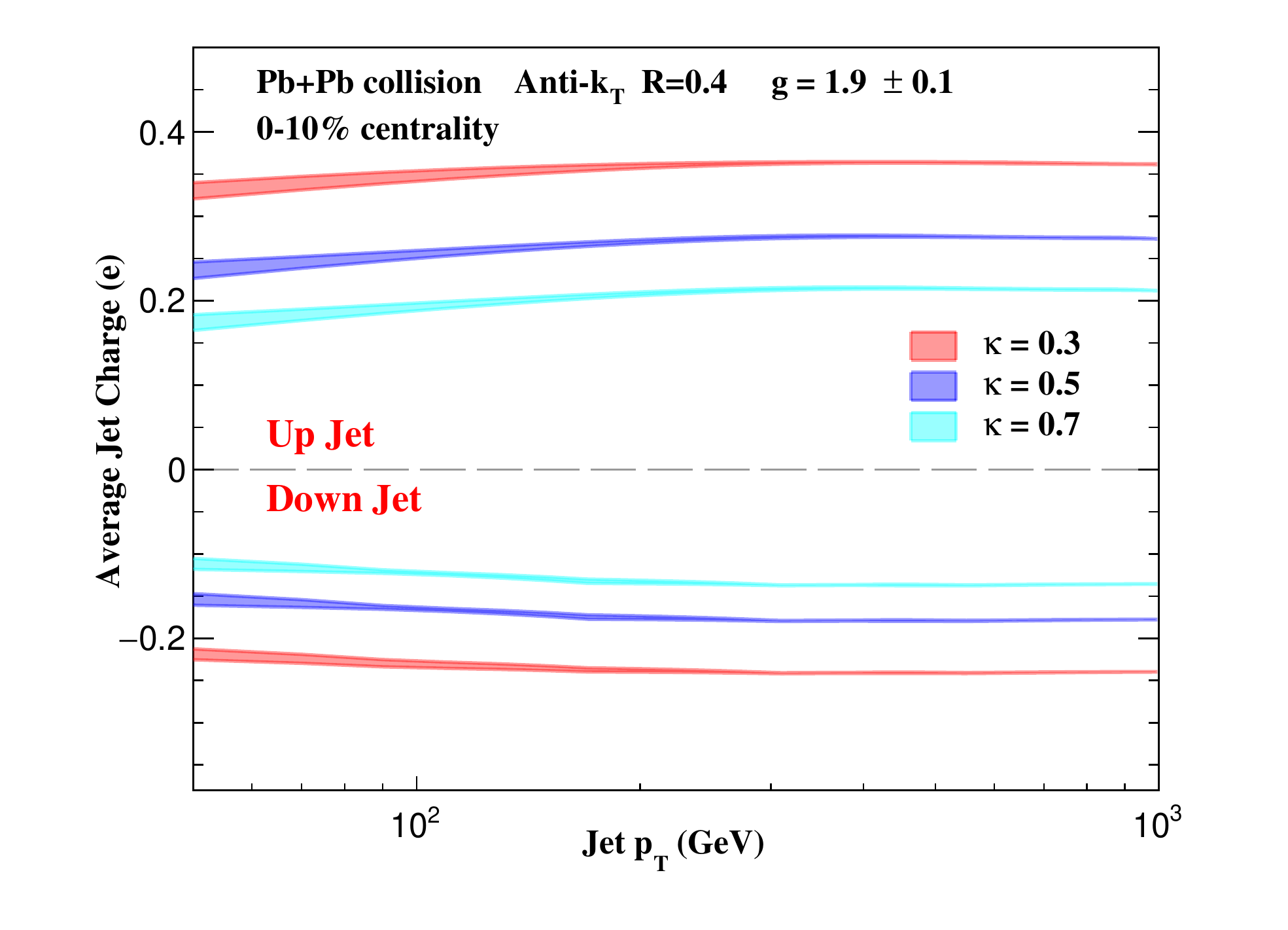}
    \vspace{-1cm}
    \caption{
    Average charges for an up-quark jet and a down quark jet as a function of jet $p_T$ in  $\sqrt{s_{\rm NN}}=5.02$ TeV Pb+Pb collisions at the LHC with
    $\kappa=$0.3, 0.5 and 0.7, respectively. 
    The coupling between the jet and QCD medium is set to be $g$=1.9$\pm$0.1. 
    }
    \label{fig:AAQ}
\end{figure}

\begin{figure}[htb!]
    \centering
  \hspace*{-.8cm}  \includegraphics[scale=0.51]{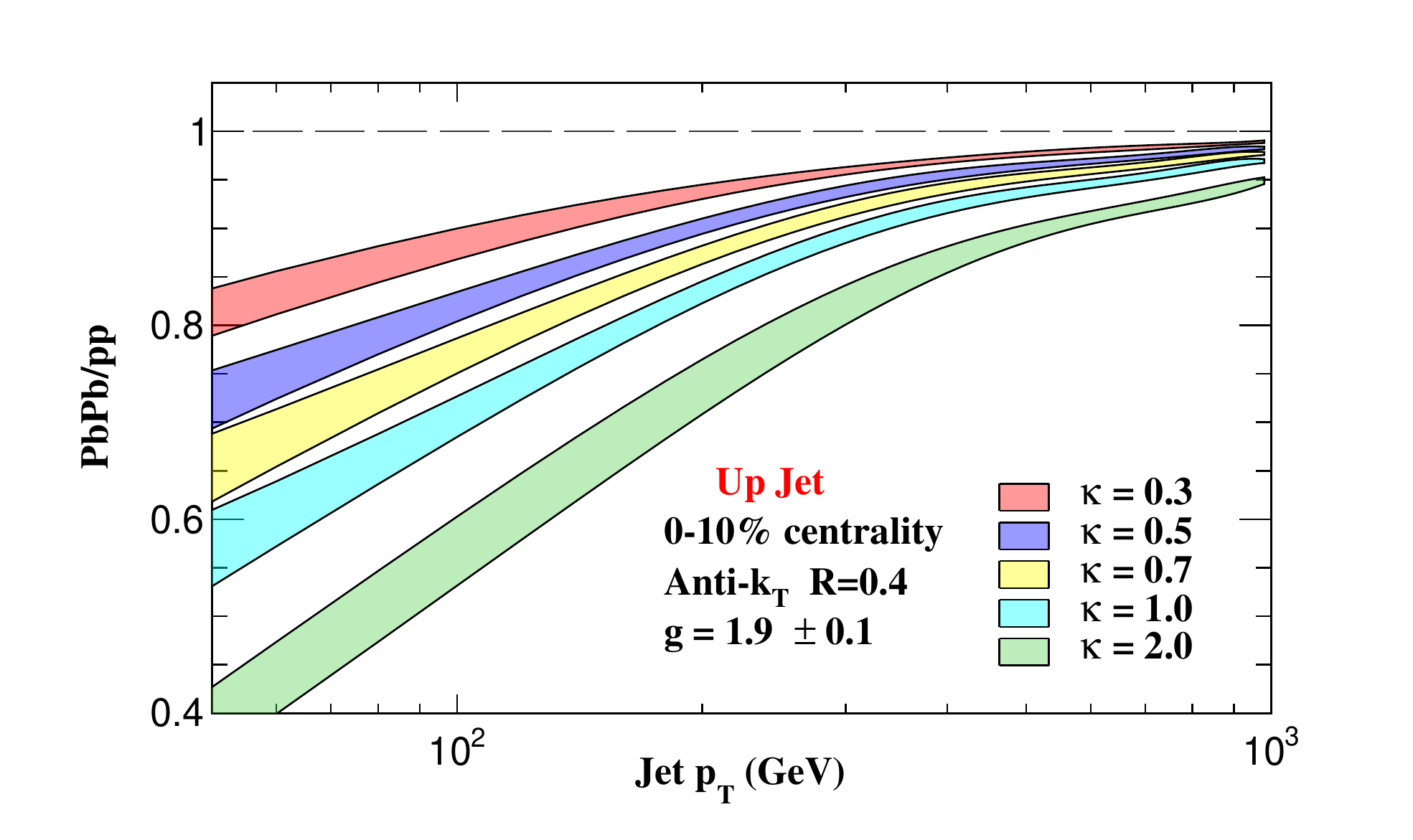}
    \vspace{-0.8cm}
    \caption{
    Modifications of the average charge for an up-quark jet as a function of jet $p_T$ in  $\sqrt{s_{\rm NN}}=5.02$~TeV 0\%-10\% central Pb+Pb collisions at the LHC.  We chose 
    $\kappa=$0.3, 0.5, 0.7, 1,  and 2, respectively.  The coupling between the jet and QCD medium is again set to be $g$=1.9$\pm$0.1. 
    }
    \label{fig:RAA}
\end{figure}

\begin{figure*}[htb!]
    \centering
    \includegraphics[scale=0.445]{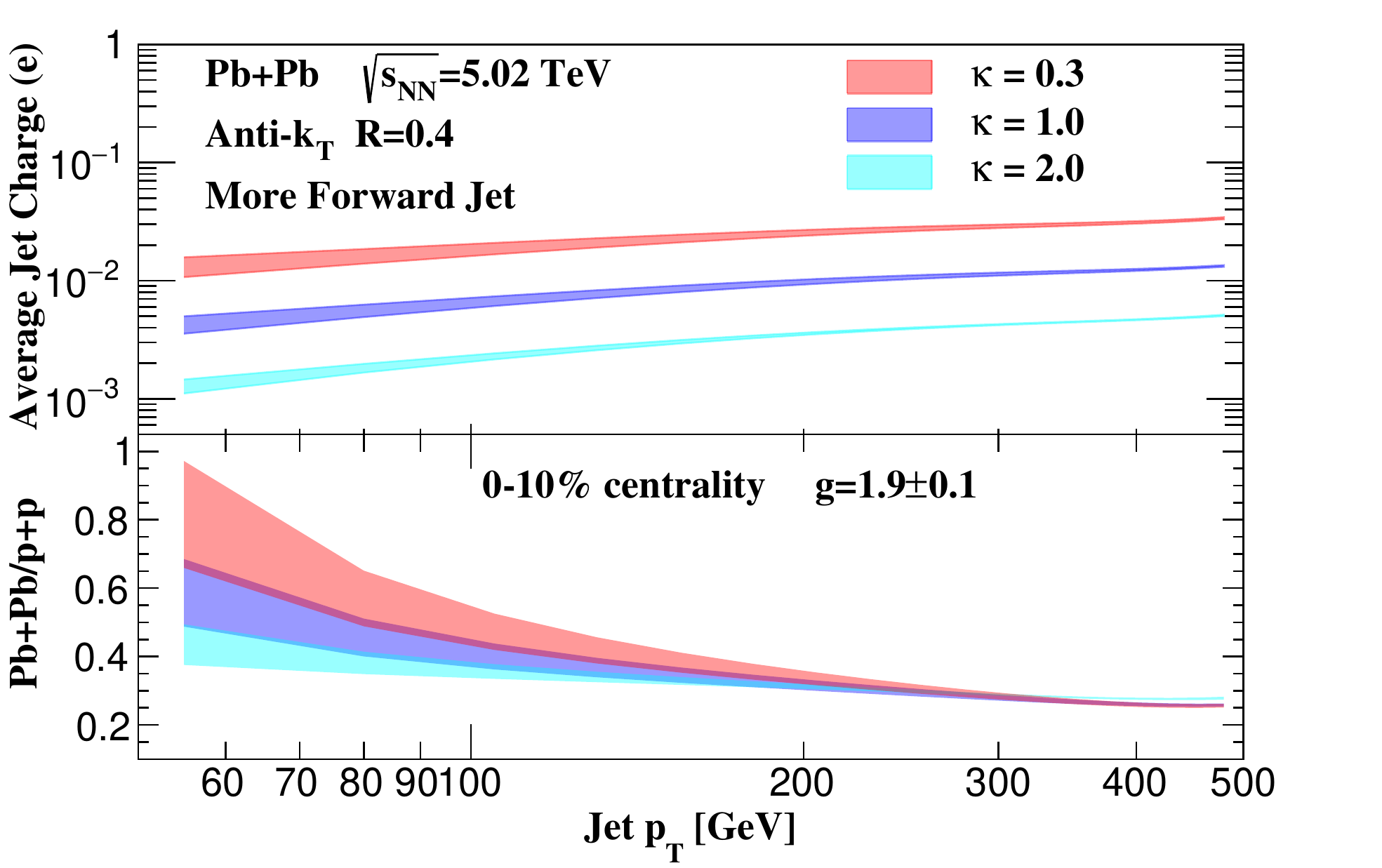}
    \includegraphics[scale=0.445]{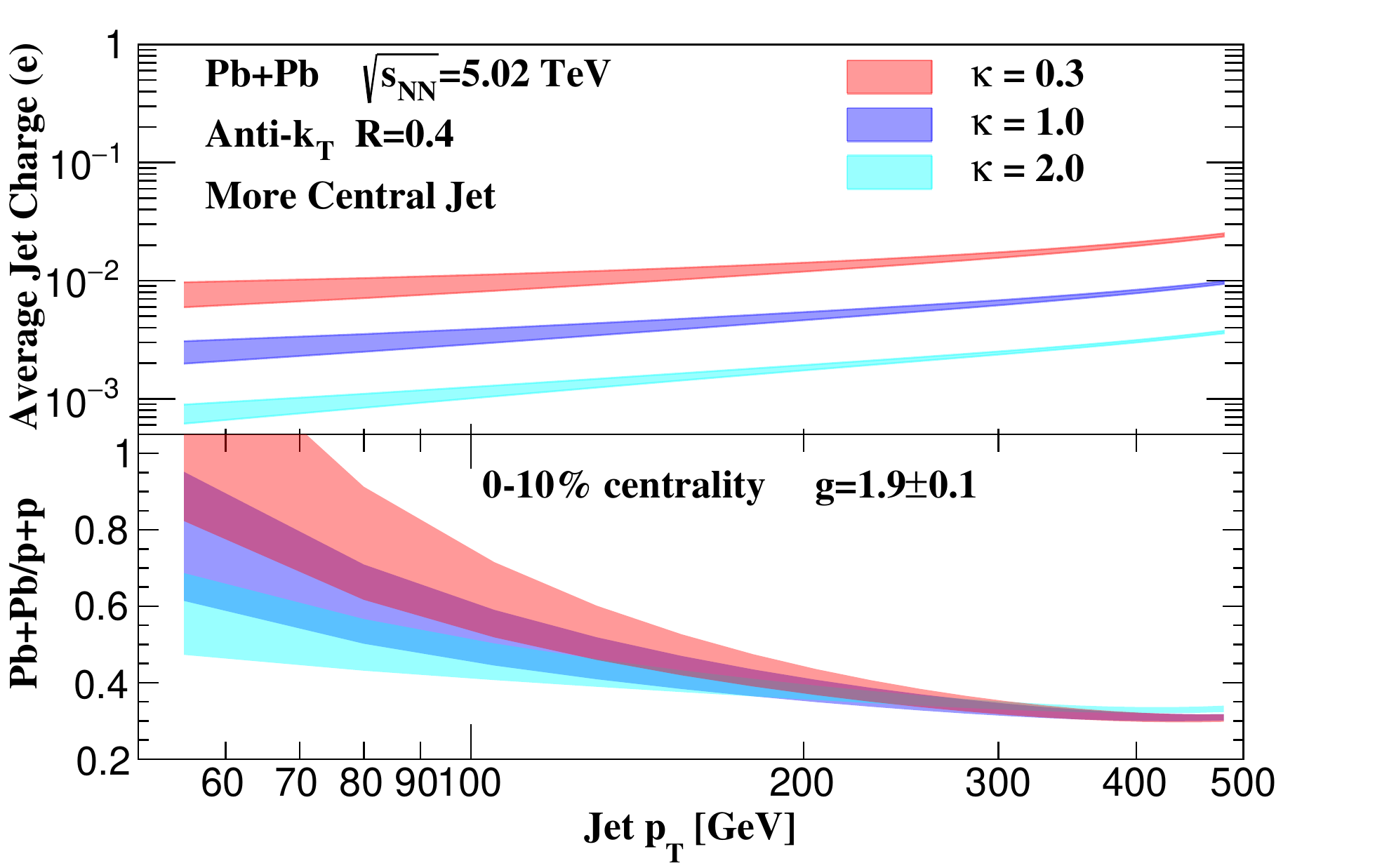}
    \vspace{-0.4cm}
    \caption{
    Average jet charge distribution for a more forward (right) and a more central (left) jet  with  $\kappa=$0.3, 1.0 and 2.0 for dijet production in Pb+Pb collisions with $\sqrt{s_{\rm NN}}=$5.02 TeV. The bottom panel shows the modification  of the average jet charge in 0\%-10\%  central Pb+Pb collisions compared with the one in  $p+p$ collisions. 
    }
    \label{fig:AA_pt}
\end{figure*}

\section{Numerical results}\label{sec:numerical}

In this section we present the numerical results for  the jet charge distribution in $p+p$ and Pb+Pb collisions. 
The average jet charge depends on the nonperturbative  parameters which are the boundary conditions of Eqs.~(\ref{eq:Qevol}) and~(\ref{eq:AAQevol}).  
In $p+p$ collisions we set the initial scale in Eq.~(\ref{eq:Qevol}) to $\mu_0=1$ GeV and evolve the fragmentation functions from $\mu_0=1$ GeV to the jet scale. 
In heavy-ion collisions, we will use the same nonperturbative parameters as in $p+p$ collisions. The medium corrections are introduced through the evolution of the fragmentation function  from $\Lambda_{\rm QCD}=0.2$ GeV to the jet scale using  in-medium splitting kernels.   In practice, according to Eq.~(\ref{eq:AAQevol}) the fragmentation function is evolved from  
$\Lambda_{\rm QCD}=0.2$ GeV, where the vacuum splitting function (evolution) is set to zero when $\mu < 1$ GeV.
This is possible since the splitting kernel is regulated by the thermal parton mass in the medium $\sim \mu_D$. 
The scale of $\alpha_s$ in the medium-induced splitting functions is chosen to be $\sqrt{k_\perp^2+\mu_D^2}$, where we take
 $\mu_D= 0.75$ GeV as an average Debye mass. We use a quark-gluon plasma medium with $N_f =2$ active quark flavors and $\mu_D = gT\sqrt{1+N_f/6} $. For $g$=1.9 the quoted average Debye mass  corresponds to an average temperature of $T \approx 340$~MeV. 
 The default jet scale is  $\mu=2E\tan R/2 \approx p_T R $. 
The fractions of different jet types are generated using LO matrix elements with CT14NLO~\cite{Dulat:2015mca} for $p+p$ collisions and nCTEQ15~\cite{Kovarik:2015cma} for Pb+Pb collisions.  For the latter case the energy loss approach is applied to calculate the modification of the fractions in the QCD medium~\cite{Kang:2014xsa}. 

Figure~\ref{fig:ppQ} shows the up- and down-quark jet charges as a function of jet $p_T$ in p+p collisions with $\sqrt{s_{\rm NN}}=8$~TeV. The average jet charge  relies on only one nonperturbative parameter/boundary condition for a given $\kappa$ and the jet type, which is obtained through PYTHIA8~\cite{Sjostrand:2007gs,Sjostrand:2014zea} simulations.  The uncertainties are evaluated by varying the factorization scale $\mu$  in Eq.~(\ref{eq:Q}) by a factor of two, and we find  that these  scale uncertainties are small. The average jet charges for up- and down-quark jets are well separated and the signs of the parent parton's charges are consistent with the ones of the jet charge. This property can be used for flavor discrimination and  quark/antiquark jet discrimination.  The absolute value of the jet charge decreases with $\kappa$, as expected from the definition in Eq.~(\ref{eq:charge}).  The predictions agree very well with the measurements by ATLAS~\cite{Aad:2015cua}, even though the data have large experimental uncertainties.   

In Fig.~\ref{fig:QoverPT}  we present the average  jet charge distribution for the more forward jet and more central jet with  $\kappa=$0.3, 0.5 and 0.7 in $p+p$ collisions. 
The dijet events are selected with kinematic cuts $p_T>50$ GeV, $|\eta|>2.1$ for both of the jets. The more forward and more central jets are the ones with a larger and smaller absolute value of rapidity in the dijet production process, respectively. For the back-to-back dijet production, more details about the kinematic cuts, which are used to calculate the fraction of quark and gluon jets, are  given in Appendix~\ref{app:fraction}. 
Assuming that the sea quark contribution to the averaged charge is roughly zero, the charge of more forward or central jets is approximately 
\begin{align}\label{eq:Qjet}
    \langle Q^{f/c}_{\kappa} \rangle = (f^{f/c}_{u}-f^{f/c}_{\bar{u}})  \langle Q_{\kappa}^{ u} \rangle+(f^{f/c}_{d}-f^{f/c}_{\bar{d}})  \langle Q_{\kappa}^{ d} \rangle \,,
\end{align}
where $f^{f/c}_q$ is the fraction of $q$ jet for the more forward or central jets and $ Q_{\kappa, q_i}$ is the average charge for $q_i$ jet.  Given the quark jet fractions and the measurements in Fig.~\ref{fig:QoverPT}, the charge of up- and down-quark jets can be extracted, which has been done by the ATLAS Collaboration~\cite{Aad:2015cua} in $p+p$ collisions, as shown in Fig.~\ref{fig:ppQ}. 
Note that  the difference between the more forward and central jets is the different fractions of quark  and gluon jets due to the different parton densities at different values of Bjorken x. Therefore,  initial-state effects are  predominantly highlighted in comparing the charges of more forward and more central jets. 
The valence quark contribution to the jet is enhanced for the more forward jet, especially in the high transverse momentum region. As a result, the average jet charge for the more forward jet is larger. The shapes in Fig.~\ref{fig:QoverPT} are mostly determined by the $p_T$ distributions of the jet flavor fractions. The overall predictions agree well with the available experimental data from Ref.~\cite{Aad:2015cua}. The predictions can be improved with a more precise calculation of the fractions, which is beyond the scope of this paper as we primarily aim to calculate the jet charge in ultrarelativistic 
nuclear collisions.

In heavy-ion collisions, we define the nonperturbative factor  as  $\tilde{D}_q^{Q,{\rm full}}(\kappa,\mu_0=\Lambda_{\rm QCD})=\tilde{D}_q^{Q}(\kappa,\mu_0=1~{\rm GeV})$, where in a QCD medium the vacuum splitting function in Eq.~(\ref{eq:AAQevol}) is set to be zero when $\mu<$ 1 GeV as discussed before. The medium modified jet charge is calculated from Eq.~(\ref{eq:AAQ}). 
Figure~\ref{fig:AAQ} shows the average up- and down-quark jet charge  with $\sqrt{s_{\rm NN}}=5.02$ TeV and  $\kappa=$0.3, 0.5 and 0.7 in 0\%-10\% central Pb+Pb collisions. The uncertainties are calculated by varying $g$, the coupling between the jet and QCD medium, in the range (1.8, 2.0).  The most important message in this figure is that  in spite of the in-medium modification the charges of up-quark jets and down-quark jets remain well separated. Thus, measurements in 
different kinematic ranges, such as the  ones carried out by the ATLAS Collaboration, hold the promise of extracting the individual flavor jet charges in analogy to  the simpler $p+p$ reactions.

This brings us to an important proposed measurement that we present in Fig.~\ref{fig:RAA}  -- the modification of individual flavor jet charges in heavy-ion versus proton collisions.      As an example, we show the medium modifications to the up-quark jet charge.  Because the only difference between the up- and down-quark jet charges is the nonperturbative parameters or boundary conditions, the modifications of the down-quark jet can be obtained through  
\begin{align}
    \frac{\langle Q_{\kappa, u}^{ \rm Pb+Pb}(p_T)\rangle }{\langle Q_{\kappa, u}^{p+p}(p_T)\rangle} =\frac{\langle Q_{\kappa,d}^{ \rm Pb+Pb}(p_T)\rangle }{\langle Q_{\kappa, d}^{p+p}(p_T)\rangle} \; . 
\end{align} 
The importance of this observable is that it eliminates the initial-state isospin effects and helps reveal the effects of the final-state medium-induced parton shower on the
jet functions and fragmentation function evolution. Thus, it is not surprising that the
medium corrections are larger for  smaller energy jets where the medium-induced splitting functions are more important. When $\kappa$ is large the $(\kappa+1)$th Mellin moment of the medium splitting function is more sensitive to the soft-gluon emissions.  In the QCD medium  jets tend  to radiate more soft gluons, in comparison to the vacuum.  As a result, as shown in Fig.~\ref{fig:RAA}, the modification is larger with a large $\kappa$. This is illustrated by the inclusion of numerical results for $\kappa = 1,\, 2$.  As discussed in Eq.~(\ref{eq:mediumQT}), the measurements of average up-quark or down-quark jets can be used to study the $(\kappa+1)$th Mellin moment of the medium-induced splitting function. By comparing  Fig.~\ref{fig:AAQ} to Fig.~\ref{fig:RAA} we see that there is  a trade-off between the  increased sensitivity to the in-medium modification of the individual jet charge and its absolute value.  An alternative way to largely eliminate initial-state isospin effects, based on studying the average jet charge in central-to-peripheral  nucleus-nucleus collisions, was discussed in Ref.~\cite{Chen:2019gqo}. The modification will then be driven by the different energy loss of quark and gluon jets.

The measurement of individual flavor jet charges will require, without a doubt, excellent statistics, experimental advances, and innovation. The average 
jet charge for the more forward and central jets 
should be  measured  relatively straightforwardly in heavy-ion collisions. Figure~\ref{fig:AA_pt}  presents our theoretical predictions for the average jet charge for the more forward and central jets in Pb+Pb collision with $\sqrt{s_{\rm NN}}=5.02$ TeV  with the kinematic cuts shown in Appendix~\ref{app:fraction}.   As indicated by Eq.~(\ref{eq:Qjet}), the medium corrections are introduced  from the modifications of the fraction of quark jets and the modifications of the average charge of the jet. Because of the existence of neutrons in the heavy nucleus, the fraction of up- and down-quark jets is significantly changed, which leads to a large modification of the jet charge in heavy-ion collisions.  A comparison between the fractions in $p+p$ and Pb+Pb collision for the more forward and central jets can be found in Appendix~\ref{app:fraction}.    For  very large $ p_T $ jets  initial state  effects are  most important, the ratio between Pb+Pb and p+p collisions is almost independent on $\kappa$.  At moderate and low $p_T$ the effect of in-medium parton showers
on the jet charge also plays a role. Our numerical results are given for values of $\kappa = 0.3, \, 1,\, 2$ and show clear sensitivity to medium-induced parton shower effects at jet transverse momenta
under 200~GeV. We conclude that measurements  of this observable over a wide kinematic range can provide insight into the interplay of
initial-state  and final-state effects in ultrarelativistic nucleus-nucleus collisions.

\section{Conclusion}\label{sec:concl}

In summary, we  developed a theoretical framework to evaluate the jet charge distributions in heavy-ion collisions.  Our work  builds upon  the SCET approach,  where the jet charge observable can be factorized into perturbatively calculable jet functions, perturbative evolution equations, and the nonperturbative fragmentation functions.  This factorization formula was validated phenomenologically through  comparison between theory and recent  measurements of the jet charge distributions in $p+p$ collisions at the LHC. 

In heavy-ion collisions, the jet functions, jet matching coefficients, and the evolution of the fragmentations are constructed with the help of the medium-induced splitting kernels derived  in the framework of SCET with Glauber gluon interactions.  Specifically, we implemented splitting kernel grids to  first order in opacity computed in a viscous hydrodynamic background to simulate QCD matter produced in heavy-ion collisions and demonstrated  how the jet charge observable can be calculated with controlled theoretical precision.  

The great utility of the jet charge observable is in the ability to discriminate between jets of various flavors, for example  up-quark jets and down-quark jets, as well as carry out  quark jet and 
antiquark jet separation. With this in mind,  we showed that the  jet charges for various flavor jets remain distinct even in the heavy-ion environment.   The modification of the jet charge  of distinct 
flavor jets  can provide novel insight into the Mellin moments of medium-induced splitting functions and the in-medium evolution of the then nonperturbative fragmentation functions. We further found that 
the moment parameter $\kappa$ in the definition of the jet charge can be used to optimize the sensitivity to the in-medium evolution effects or the magnitude of the observable. 

We further presented theoretical  predictions for the average jet charge without flavor separation for the more forward and central jets for dijet production in heavy-ion collisions, which can be 
measured at the LHC and RHIC.  For very high transverse momentum jets the nuclear modification  is dominated by an initial-state  isospin  effect, as was also found in a recent Monte Carlo study~\cite{Chen:2019gqo}.
For intermediate and small transverse momenta parton showers induced by QCD matter can play an important role in the average jet charge modification. Thus, we suggest that studies of this
observable in the kinematic range covered by the future sPHENIX experiment at RHIC can also be quite illuminating.

The jet charge definition is independent of the hard process; however, different hard processes can change significantly  the fraction of quark or gluon jets. In addition to dijet production, average jet charge can be measured in vector boson plus jet production, or heavy flavor jet production in proton and heavy-ion collisions. After this work was completed, using an inclusive  jet sample, the CMS Collaboration  presented the first measurement of  the jet charge in heavy-ion collisions~\cite{Sirunyan:2020qvi}.  This serves as motivation to evaluate such observables with higher perturbative precision, improved baseline determination, and in-medium evolution to higher orders in opacity in the future.   We finally remark that this observable can also be studied at  an electron-ion collider.

\acknowledgments{
 This work was supported by the U.S. Department of Energy under Contract No. DE-AC52-06NA25396  and the Los Alamos National Laboratory LDRD program.
}

\appendix
\counterwithin{figure}{section}

\section{Jet function and jet matching coefficients} \label{app:JetF}

We will show here how to calculate the jet matching coefficient in the vacuum ($q\to q g$ channel), with an emphasis
on a representation which is useful to define the medium corrections. 

The amplitude for the $q (p+l)\to q(p) g(l)$ splitting is given by
\begin{align}
  \frac{1}{2 N_c x \omega}  |\overline{\mathcal{M}}|^2 = \frac{C_F g^2_s}{l_\perp^2}\omega (d(1-x)^2-2(x^2-4x+1)) \; ,
\end{align}
where $d$ is the number of dimensions and   $\omega$ is the large lightcone component of the parent parton. $C_F = 4/3$ and $N_c = 3$ for  SU$_c$(3). 
The phase space integral that we need to perform with $d=4-2\epsilon$ reads 
\begin{align}
    & \left(\frac{\mu^2 e^{\gamma_E}}{4 \pi}\right)^\epsilon \int \frac{d^d l}{(2\pi)^{d-1}} \delta(l^--(1-x) \omega)\delta(l^2) 
        \nonumber \\
    &  \times 
    \delta(s-\omega(l^++p^+))  =
    \int ds \frac{e^{\gamma_E} s^{-\epsilon} x^{1-\epsilon}(1-x)^{-\epsilon}}{16 \pi^2 \Gamma(1-\epsilon) \omega} \;, 
\end{align}
where $s$ is the invariant mass of the jet 
\begin{align}
    s=w(k^++l^+) = \frac{ l_\perp^2}{x(1-x)} \; .
\end{align}

We obtain the NLO matching coefficient as follows
\begin{multline}
    \mathcal{J}_{qq}^{(1)}(E,R,x,\mu) = 
    \\
    \frac{C_F \alpha_s}{2\pi} \frac{e^{\gamma_E}}{\Gamma(1-\epsilon)}\int_0^{s_{\rm max} }
     \frac{ds}{s} \left(\frac{\mu^2 }{s}\right)^\epsilon \frac{1+x^2-\epsilon(1-x)^2}{x^\epsilon(1-x)^{1+\epsilon}} \;,
\end{multline}
where $s_{\rm max}=4x(1-x)E^2\tan(R/2)^2$  depends on the jet radius $R$. It can be also written  in the following form
\begin{multline} \label{eq:jq2q}
    \mathcal{J}_{qq}^{(1)}(E,R,x,\mu) = 
    \\
    \frac{C_F \alpha_s}{2\pi}   \frac{e^{\epsilon\gamma_E }}{\Gamma(1-\epsilon)} \int
    \frac{ d l_\perp^2}{l_\perp^2} \left(\frac{\mu^2 }{l_\perp^2}\right)^\epsilon \frac{1+x^2-\epsilon(1-x)^2}{1-x} \;, 
\end{multline}
where $0<l_\perp< 2x(1-x)E\tan(R/2)$. The utility of this representation is that the jet matching coefficients $\mathcal{J}_{qq}$ in Eq.~(\ref{eq:jq2q}) 
is expressed in the form of the integral of the splitting kernel, which is the starting point to construct the medium corrections. 
After performing the $l_\perp$ integration we obtain the same matching coefficient as derived in Refs~\cite{Waalewijn:2012sv,Jain:2011xz,Ellis:2010rwa}. 
The total jet function can be calculated from  the expression
\begin{align} \label{eq:totaljet}
    J_{q}(E,R,\mu) = \int_0^1 dz z \big[\mathcal{J}_{qq}(E,R,z,\mu)+\mathcal{J}_{qg}(E,R,z,\mu)\big] \;.
\end{align}

\section{Fractions of jets initiated by different parton flavors in p+p and Pb+Pb collisions }
\label{app:fraction}

\begin{figure*}[htb]
    \centering
    %\vspace{0.5cm}
    \includegraphics[scale=0.44]{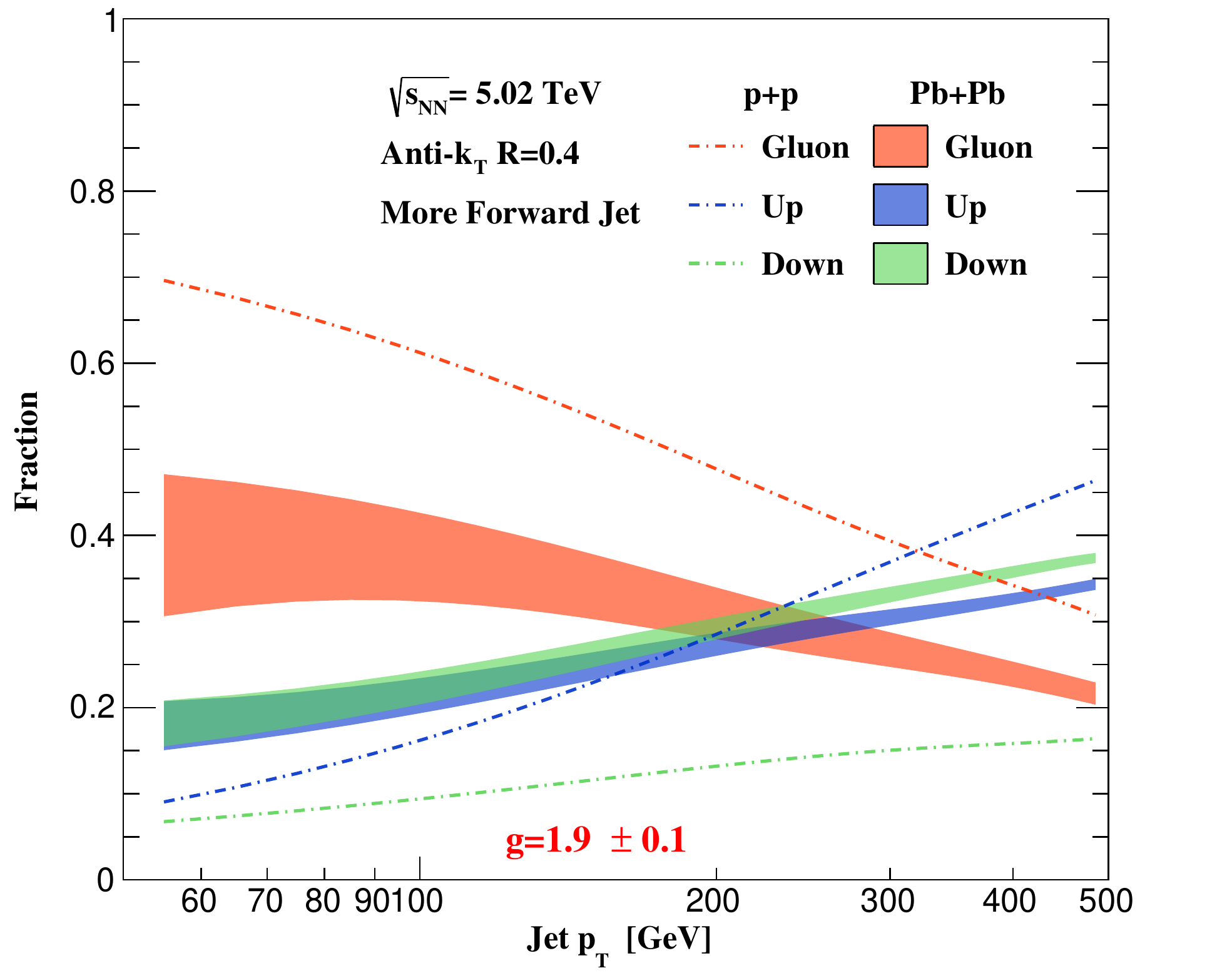}
    \includegraphics[scale=0.44]{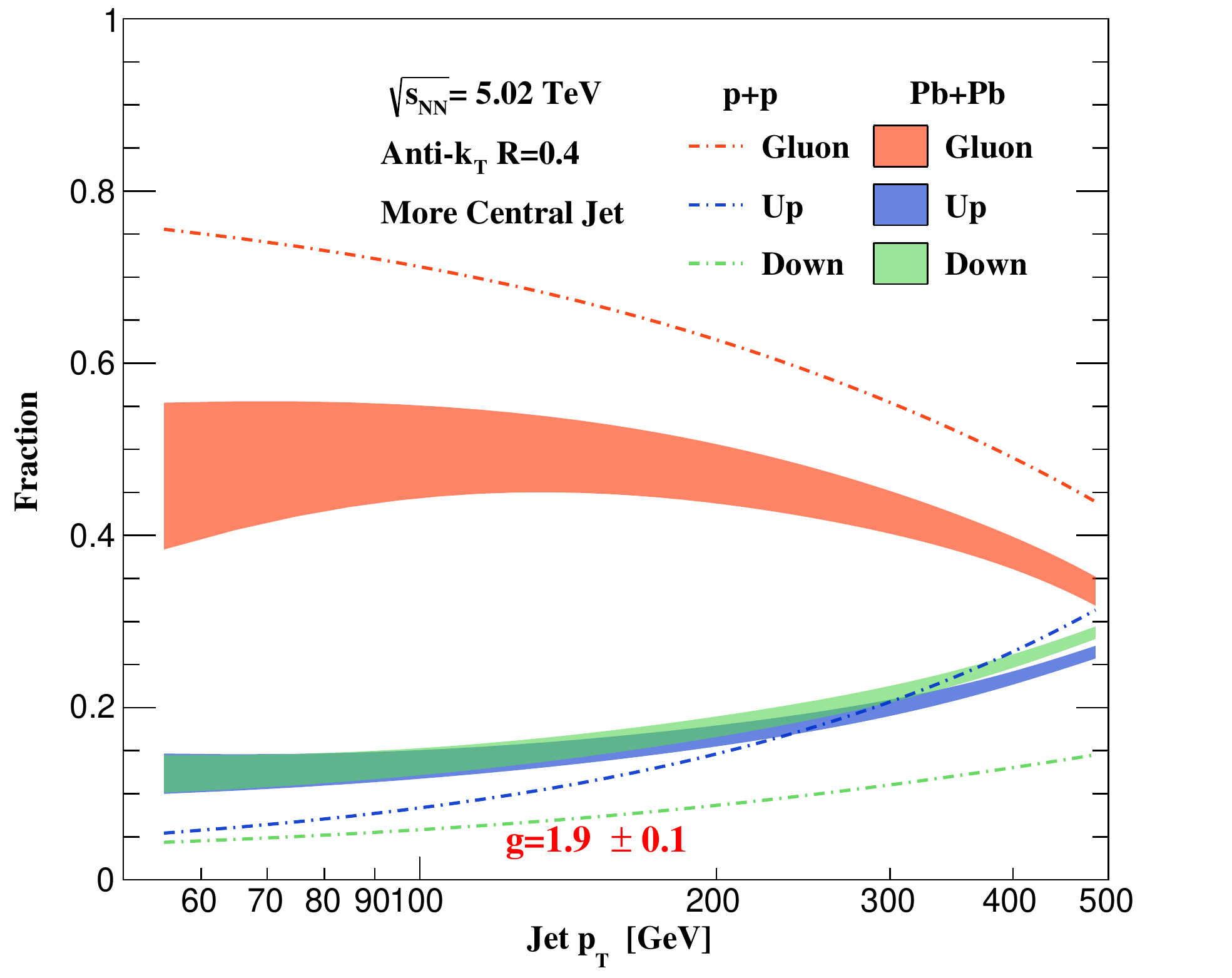}
    \vspace{-0.4cm}
    \caption{The fractions of the gluon, up-quark and down-quark jet for the case of more forward (left) and more central (right) jet in p+p and Pb+Pb collisions with $\sqrt{s_{\rm NN}}=$5.02 TeV.   }
    \label{fig:fraction}
\end{figure*}

The fractions of the different jet types are simulated using LO matrix element for dijet production with CT14NLO PDFs~\cite{Dulat:2015mca} for  proton and nCTEQ15 PDFs~\cite{Kovarik:2015cma}  for lead. The jet is reconstructed with the anti-$k_T$ $R=0.4$ algorithm. 
We choose the events with at least two jets satisfying $p_T>50$ GeV and $|\eta|<2.1$. In order to clearly match the hard-scattering quark or gluon to outgoing jet we impose a cut on the ratio of the leading and subleading jet's transverse momenta $p_T^{\rm lead}/p_T^{\rm sublead}<1.5$. The same kinematic constraints were used by the measurements~\cite{Aad:2015cua} at the LHC. 

In heavy-ion collisions in addition to the initial-state effects, jets cross sections are suppressed,  or quenched,  due to the interaction with the hot QCD medium.  We take this effect into account in the calculations that follow.
 Figure~\ref{fig:fraction} shows the fraction of gluon, up-quark and down-quark jet in p+p and Pb+Pb collisions.  The fraction of gluon jet in Pb+Pb collisions is smaller because the gluon jet tend to lose significantly more energy in QCD matter relative to quark jets. The $^{208}{\rm Pb}$ nucleus contains 82 protons and 126 neutrons, as a result when compared to  proton-proton collisions the fraction of up-quark jet is reduced while the fraction of down-quark jet is enhanced significantly.  The left panel of Figure~\ref{fig:fraction}  shows results for more forward jets, defined as the jet with a larger absolute value of rapidity in dijet production.  The right panel of Figure~\ref{fig:fraction}  shows results for more central jets, defined as  the jet with a smaller absolute value of rapidity.      The bands in Pb+Pb collisions represent the uncertainties by varying the coupling between the jet and QCD medium in the range $1.8<g<2.0$. 

\bibliography{mybibfile}

\end{document}